\PassOptionsToPackage{hyphens}{url}
\PassOptionsToPackage{hidelinks}{hyperref}
\documentclass[runningheads]{llncs}
\usepackage[T1]{fontenc}
\usepackage{hyperref}
\usepackage[type={CC},modifier={by},version={4.0},imagewidth=5em]{doclicense}
\usepackage[inline]{enumitem}
\usepackage{amsmath}
\usepackage{amssymb}
\usepackage[frozencache]{minted}
\usepackage{listings}
\usepackage{fontawesome}
\usepackage{relsize}
\usepackage{soul}

\usepackage{booktabs}
\usepackage{array}
\newcolumntype{H}{>{\setbox0=\hbox\bgroup}c<{\egroup}@{}}

\usepackage[capitalize]{cleveref}
\usepackage[svgnames]{xcolor}

\usepackage{orcidlink}

\newif\ifdraft\draftfalse
\newif\iflong\longfalse

\ifdraft
\usepackage{todonotes} %
\newcommand\pagest[1]{ (#1 pages)}
\pdfpagesattr{/CropBox [92 100 523 740]} %
\else
\usepackage[disable]{todonotes}
\newcommand\pagest[1]{}
\fi

\newcommand\name{CUTECat\xspace}
\newcommand{\synvar}[1]{\ensuremath{#1}}
\newcommand{\synkeyword}[1]{\textcolor{red!60!black}{\texttt{#1}}}
\newcommand{\synpunct}[1]{\textcolor{black!40!white}{\texttt{#1}}}

\newcommand{\synjust}{~\synpunct{:-}~}

\newcommand{\syncomma}{\synpunct{,}}

\newcommand{\syntrue}{\synkeyword{true}}
\newcommand{\synfalse}{\synkeyword{false}}

\newcommand{\synlangle}{\synpunct{$\langle$}}
\newcommand{\synrangle}{\synpunct{$\rangle$}}
\newcommand{\synmid}{\synpunct{~$|$~}}
\newcommand{\synemptydefault}{\synvar{\varnothing}}
\newcommand{\synerror}{\synvar{\circledast}}
\newcommand{\synempty}{\synemptydefault}
\newcommand{\synconflict}{\synerror}

\newcommand{\synellipsis}{\synpunct{,$\ldots$,}}

\newcommand{\syndef}{$ ::= $}
\newcommand{\synalt}{\;$|$\;}

\newcommand{\exctx}[1]{\textcolor{blue!80!black}{\ensuremath{#1}}}

\newcommand{\exeval}{\exctx{\;\longrightarrow\;}}
\newcommand{\exevalstar}{\exctx{\;\longrightarrow^*\;}}

\usepackage{stmaryrd}

\newcommand{\trueconstr}[1]{(#1)}
\newcommand{\falseconstr}[1]{\lnot(#1)}

\newcommand{\concrete}[1]{\ensuremath{\mathbf{#1}}}
\newcommand{\concreteTrue}{\concrete{true}\xspace}
\newcommand{\concreteFalse}{\concrete{false}\xspace}

\newcommand\default[3]{\ensuremath{\synlangle #1 \synmid #2 \synjust #3 \synrangle}}
\newcommand\pdefault[2]{\ensuremath{\synlangle #1 \synjust #2 \synrangle}}

\usepackage{subcaption}
\usepackage{multicol}

\newcommand\concolic[3]{\ensuremath{#1 \vdash #2 \mid #3}}
\newcommand\concolicloc[2]{\ensuremath{#1 \mid #2}}
\newcommand\concolicglo[2]{\ensuremath{#1 \vdash #2}}

\DeclareRobustCommand{\rchi}{{\mathpalette\irchi\relax}}
\newcommand{\irchi}[2]{\raisebox{\depth}{$#1\chi$}} %

\usepackage{wrapfig}
\crefname{section}{Sec.}{Secs.}
\crefname{figure}{Fig.}{Figs.}
\crefname{example}{Ex.}{Exs.}
\crefname{theorem}{Th.}{Ths.}

\BeforeBeginEnvironment{minted}{\vspace{-.25cm}}
\AfterEndEnvironment{minted}{\vspace{-.5cm}}
\setminted{fontsize=\scriptsize}
\setmintedinline{fontsize=\footnotesize}

\usepackage{enumitem}
\setlist[itemize]{leftmargin=*,noitemsep, topsep=0pt}
\setlist[description]{leftmargin=*,noitemsep, topsep=0pt}

\usepackage[numbers,sort&compress]{natbib}
\usepackage{mathpartir}

\def\dsymb#1{\;\,{\stackrel{\mbox{\tiny def}}{#1}}\;\,}
\def\deq{\dsymb{=}}

\title{\name: Concolic Execution\\ for Computational Law}

\newif\ifreview\reviewfalse
\ifreview
  \author{}
  \institute{}
  \usepackage{lineno}
  \linenumbers
\else
\author{Pierre Goutagny\inst{1}\orcidlink{0000-0003-0876-7188} \and
  Aymeric Fromherz\inst{2}\orcidlink{0000-0003-2642-543X} \and
  Raphaël Monat\inst{1}\orcidlink{0000-0001-8487-0326}
  }
\authorrunning{P. Goutagny, A. Fromherz and R. Monat}
\institute{
  Univ. Lille, Inria, CNRS, Centrale Lille, UMR 9189 CRIStAL, F-59000 Lille, France\\
  \and Inria Paris, France
  \email{\{pierre.goutagny,aymeric.fromherz,raphael.monat\}@inria.fr}}
\fi

\newcommand\todolater[2][]{\todo[color=Plum,#1]{#2}}

\begin{document}

\maketitle

\begin{abstract}
Many legal computations, including the amount of tax owed by a citizen, whether
they are eligible to social benefits, or the wages due to civil state servants,
are specified by \emph{computational laws}.
Their application, however, is performed by expert computer programs intended
to faithfully transcribe the law into computer code. Bugs in these programs
can lead to dramatic societal impact, e.g., paying employees incorrect amounts,
or not awarding benefits to families in need.

To address this issue, we consider concolic unit testing, a combination of
concrete execution with SMT-based symbolic execution, and propose \name, a
concolic execution tool targeting implementations of computational laws. Such
laws typically follow a pattern where a base case is later refined by many
exceptions in following law articles, a pattern that can be formally modeled
using \emph{default logic}. We show how to handle default logic inside a
concolic execution tool, and implement our approach in the context of Catala, a
recent domain-specific language tailored to implement computational laws. We
evaluate \name on several programs, including the Catala implementation of the
French housing benefits and Section 132 of the US tax code. We show that \name
can successfully generate hundreds of thousands of testcases covering all
branches of these bodies of law. Through several heuristics, we improve \name's
scalability and usability, making the testcases understandable by lawyers and
programmers alike. We believe \name paves the way for the use of formal
methods during legislative processes.
\end{abstract}

\section{Introduction}

Since at least the Sumerian empire and the code of
Ur-Nammu~\cite{ur-nammu}, human societies have been governed by laws.
From constitutional law to environmental law, including criminal law,
immigration law or intellectual property law, laws are applied
in a wide range of contexts, representative of the diverse facets of
modern societies.
Laws are typically stated in natural language, e.g., English, and therefore
require human interpretation to determine when and how they must apply: to do
so, a criminal trial might rely on a grand jury to determine the guilt of a
defendant, while companies typically hire highly specialized lawyers to ensure
that a merger is performed according to corporate law.

Other laws, however, focus on defining well-specified computations, for instance,
the amount of taxes that a household owes, the wages of civil state servants,
or whether a given family is eligible to social benefits depending on their
situation. As they must be applied to a large number of citizens, such laws,
commonly known as \emph{computational laws}, are typically implemented in
expert computer systems, which are able to automate the computation for
any input.

Unfortunately, as for any computer program, legal expert systems are not immune
to bugs, which can have tremendous consequences, either for the states or their
citizens.  Case in point, more than half of all Canadian civil servants
suffered pay issues from the Phoenix automated payroll system, resulting in
years of fixing incurred financial issues for civil servants, and, for the
Canadian government ``\$2.2 billion in unplanned expenditures'', while the
program should have provided ``\$70 million in annual savings by centralizing
pay operations'' \cite{mockler2018phoenix}.
Similarly, bugs in Louvois, the French army payroll system, led to
several years of incorrect payments to military households, either
causing missed wages or overpayments that had to be reimbursed by
individuals years later~\cite{louvois}.
Such cases are not isolated incidents, and examples of
public legal systems that were either faulty or abandoned during
development abound in many countries~\cite{redden2022automating}.

One of the core issues lies with the particular structure of computational law,
commonly consisting of a base case refined by exceptions spread out throughout
law texts. This structure, corresponding to \emph{default
logic}~\cite{brewka2000prioritizing,lawsky2017logic}, is not straightforward to
encode using modern programming languages. This leads to discrepancies between
code and law, where design choices made by programmers can be hard to
understand by lawyers.  Conversely, maintaining such implementations when
legislative processes modify the law becomes tricky, as programmers cannot
necessarily accurately pinpoint required changes in their code. %

To address this issue, \citet{merigoux2021catala} recently introduced Catala, a
domain-specific language (DSL) tailored to implement computational laws. At the
heart of Catala lies programming constructs called \emph{default terms}, which
faithfully implement default logic. Default terms are first-class citizens
in the core Catala language. To simplify their use, they are exposed
through a custom syntax designed to be understandable by programmers and lawyers alike.
This syntax enables Catala programmers to define the base case of a default term
and its corresponding exceptions in different parts of the code, thus accurately
reflecting the structure of legal texts. By combining
this feature with literate programming, Catala enables a programming style
where an official legal text is intertwined with its implementation, thus
allowing programmers to work in concert with lawyers to ensure a faithful
translation into code of the law.

To minimize common programming issues, Catala's design purposely avoids risky
programming languages constructs, e.g., by providing infinite-precision numbers
rather than floating-point ones, and by avoiding NULL values. Despite this
effort, legal implementations in Catala can nevertheless contain bugs for three
reasons. First, by heavily operating on numerical values such as amounts of money,
Catala is not immune to standard runtime errors such as divisions by zero,
possibly leading to crashes. Second, in spite of its peer-programming approach
where lawyers and programmers work together to implement the law, translations
of legal texts into code amount to legal interpretations, which can be erroneous.
Last, the law itself might exhibit inconsistencies, for instance, by defining
contradicting exceptions under specific circumstances, or by forgetting to consider
some situations -- in Catala implementations, this would lead to
a runtime exception being raised by the program.

To circumvent these issues, public administrations heavily rely on testing; some
departments are in charge of manually handcrafting testcases, interpreting the
law to compute results, and comparing them to the outputs of legal
expert systems to ensure their conformance with legal texts. Unfortunately, the
high number of cases, and the regular modifications of laws make this manual
process costly and painstaking, and do not guarantee that corner cases are not
missed.

To address this problem, this paper therefore advocates applying formal
verification techniques to computational law. Doing so raises several
challenges. First, formal verification tools must be able to efficiently
reason about the core concept of computational law, namely, default logic.
Second, to facilitate its adoption, the verification process must not
require expert knowledge of formal verification, and thus needs to be
as automated as possible. Last but not least, when identifying issues, it is crucial that
verification tools provide concrete examples usable by lawyers, so that they
can compare them with legal texts and determine whether these correspond to
law inconsistencies.

Given these constraints, we focus our attention on the application of
concolic unit
testing~\cite{DBLP:conf/sigsoft/SenMA05,DBLP:conf/pldi/GodefroidKS05}, a
combination of concrete execution with SMT-based symbolic execution, which we
believe to hit a sweet spot when it comes to reasoning about computational law.
Computational law, and thus its implementation in Catala, is deterministic, and
does not contain loops or recursion, alleviating a well-known challenge for
concolic testing~\cite{cadar13symbolic,godefroid07compositional}.
Additionally, its automated nature and the concrete inputs
it generates during analysis pave the way for use by lawyers and programmers
alike.

Our contributions are the following: we show how to concolically execute
computational law programs by providing a formal concolic semantics for default
terms (\Cref{sec:default:concolic}). Relying on this formal model, we then
implement \name, a concolic execution execution engine for Catala programs
(\Cref{sec:overview}). \name aims to detect all runtime errors, such as
divisions by zero but also law inconsistencies due to missing, or conflicting
interpretations. \name includes several optimizations, both
to improve its performance and scalability, as well as its usability, aiming to
generate human-friendly testcases to facilitate legal interpretation by lawyers
(\Cref{sec:optims}). Finally, we experimentally evaluate \name on a range of
Catala programs (\Cref{sec:evalexp}). Relying on real-world codebases from
French and US laws, we first perform an ablation study to evaluate the impact of
our different optimizations. We then conclude by empirically demonstrating that
\name can scale to the largest real-world Catala codebase currently available,
namely, the implementation of the French housing benefits (19655 lines of
Catala, including specification), generating 186390 tests in less than 7 hours of CPU time.
Our development is open-source and publicly available on GitHub; all experimental claims made in this paper are documented in an artifact
\cite{cutecat_esop25_artifact}.

\section{Encoding Default Logic in Concolic Execution}
\label{sec:encodingdefault}

In this section, we present our encoding of default logic into concolic execution.
We start with some background about computational law, and how it relates to default logic.

\subsection{Computational Law and Default Logic}
\label{sec:defaultterms}

Throughout this paper, we will rely on a simplified income tax computation as a
running example.  This computation determines the amount of taxes a household
must pay, depending on their income.  Typically, regulations specify a tax
rate, that is multiplied by the household income to determine the income tax to
pay. For instance, if the tax rate is set to 20\%, a household earning
\$100,000 would pay \$20,000 in income tax. In practice, tax laws commonly
define different tax rates for different income brackets; we omit this in our
example for simplicity.

To address several societal issues, this computation can however be modified
depending on households. For instance, low-income households (which we define in
this example as earning no more than \$10,000) might be taxed to a reduced rate
of 10\%. Furthermore, large households (which we define in this example as
households with three or more children), might benefit from a tax break, for
instance, lowering their tax rate to 15\%. For now, we do not set any
interpretation priority for households having both a low income and a large
number of children.

This structure is typical of computational laws: we have one \emph{base case}
(the tax rate is 20\%), followed by several \emph{conditional exceptions},
capturing specific situations (large, or low-income households).  Implementing
this structure in traditional programming languages is challenging for two
reasons. First, exceptions in legal texts are commonly spread out through
several law articles, which does not match the standard monolithic variable or
function definitions. Second, legal implementations should closely follow the
structure of the law they implement. Implementations correspond to applications of
the law; any difference with the original text thus amounts to a
legal reinterpretation, which should be performed by lawyers, and not
programmers. Heavily transforming the law to match traditional programming
idioms is therefore ill-advised.
To address this issue, earlier work proposed \emph{default
terms}~\cite{merigoux2021catala}, heavily inspired by default
logic~\cite{brewka2000prioritizing,lawsky2017logic}.

Default terms take the form $\default{\synvar{e_1}, \ldots,
  \synvar{e_n}}{\synvar{e_{just}}}{\synvar{e_{cons}}}$, where $\synvar{e_1},
\ldots, \synvar{e_n}$ are default expressions called `exceptions',
$\synvar{e_{just}}$ is a boolean expression called `default condition', and
$\synvar{e_{cons}}$ is a default expression. Empty values are written as
$\synemptydefault$, and conflict values, when two exceptions occur at the same
time, as $\synerror$. Due to typing guarantees, omitted here for brevity, the
default condition never reduces to $\synemptydefault$ or $\synerror$. The
grammar and formal semantics of default terms, adapted from
\cite{merigoux2021catala}, are available in \Cref{fig:grammar} and
\Cref{fig:default-reduction} respectively. Expressions and values also contain
numeric and boolean expressions, whose semantics is standard and thus omitted
from the presentation.  According to this semantics, default terms are executed
as follows:

\begin{figure}[!t]
  \centering
  \begin{center}
    \begin{tabular}{llrlr}
      Value         & \synvar{v} & \syndef & \syntrue \synalt \synfalse \synalt \synvar{n} & boolean, integer literals\\
                    &            & \synalt & \synemptydefault & empty value\\
                    &            & \synalt & \synerror & conflict value\\

      \\ %
      Expression    & \synvar{e} & \syndef & \synvar{x}\synalt\synvar{v}  & variable, values  \\
                    &            & \synalt & \synvar{e} \ensuremath{\bowtie} \synvar{e} & integer, boolean binary operators\\
                    &            & \synalt & \synlangle $\overrightarrow{\synvar{e}} \synmid\synvar{e}\synjust\synvar{e}$\synrangle                                  & default term        \\
    \end{tabular}
  \end{center}

  \caption{Grammar of default terms}
  \label{fig:grammar}
\end{figure}

\begin{figure}[t]
  \centering
  \begin{mathpar}
    \inferrule[DefaultExpr]
    {e \exevalstar v\quad\quad\quad\quad v \neq \synerror{}}
    {\synlangle v_1 \synellipsis v_i , e \synellipsis \synmid e_{just}  \synjust e_{cons} \synrangle
      \exeval \synlangle v_1 \synellipsis v_i, v \synellipsis \synmid e_{just} \synjust e_{cons} \synrangle}

    \inferrule[DefaultError]
    {e \exevalstar \synerror{}}
    {\synlangle v_1 \synellipsis v_i, e \synellipsis \synmid e_{just} \synjust e_{cons} \synrangle \exeval \synerror{}}

    \inferrule[DefaultTrueNoExceptions]
    {e_{just} \exevalstar \syntrue}
    {\synlangle \synemptydefault{}\synellipsis\synemptydefault{}\synmid e_{just} \synjust e_{cons} \synrangle\exeval e_{cons}}

    \inferrule[DefaultFalseNoExceptions]
    {e_{just} \exevalstar \synfalse}
    {\synlangle \synemptydefault{}\synellipsis\synemptydefault{}\synmid e_{just} \synjust e_{cons} \synrangle\exeval \synemptydefault{}}

    \inferrule[DefaultOneException]
    {v \neq \synemptydefault, \synerror}
    {\synlangle \synemptydefault\synellipsis\synemptydefault\syncomma v \syncomma\synemptydefault\synellipsis\synemptydefault
      \synmid  e_{just} \synjust e_{cons}
      \synrangle\exeval v}

    \inferrule[DefaultExceptionsConflict]
    {v_i \neq \synemptydefault \\ v_j \neq \synemptydefault \\ \forall k, v_k \neq \synerror}
    {\synlangle v_1 \synellipsis v_i \synellipsis v_j \synellipsis v_n \synmid
      e_{just} \synjust e_{cons} \synrangle\exeval \synerror{}}
  \end{mathpar}

  \caption{Selected reduction rules for default terms}
  \label{fig:default-reduction}
\end{figure}

\begin{itemize}
\item If all exceptions reduce to empty values, that is, if no exception is
raised or if there are none, then the default expression reduces either to
\synvar{e_{cons}} if condition \synvar{e_{just}} evaluates to true
(\textsc{DefaultTrueNoExceptions}), or to \synemptydefault{} otherwise
(\textsc{DefaultFalseNoExceptions}). For instance, the term
\default{}{\synfalse}{1} reduces to \synemptydefault.
\item If exactly one exception \synvar{e_i} reduces to a non-empty value,
then the default expression reduces to its value (\textsc{DefaultOneException}).
Therefore, the term
\default{\default{}{\syntrue}{1},\default{}{\synfalse}{2}}{\syntrue}{3}
reduces to 1.
\item If more than one exception expression reduces to a non-empty value, that
is if several exceptions are raised at the same time, then the default
expression reduces to a conflict error \synerror{}
(\textsc{DefaultExceptionsConflict})%
        \footnote{Two raised exceptions are considered a conflict even if they lead to the same result.}%
        . Thus,
\default{\default{}{\syntrue}{1},\default{}{\syntrue}{2}}{\syntrue}{3} yields
\synerror.
\end{itemize}

\begin{remark}
  \label{not:defaults}
  To lighten notations, we will omit the list of exceptions from a default term when
it is empty. For instance, the term $\default{}{e_1}{e_2}$ will be written as
$\pdefault{e_1}{e_2}$.
\end{remark}

Encoding the income tax rate we described earlier as a default term is
then straightforward. As the base case always applies, the default condition
$e_{just}$ is trivial, and the two cases for low-income and large households
are encoded as exceptions, leading to the following term:

\newcommand\runningdefault{\ensuremath{
  \default
  {\pdefault{\texttt{income} \leq \$10,000}{10\%}, \pdefault{\texttt{nb\_children} \ge 3}{15\%}}
  {\syntrue}
  {20\%}
  }
}
\medskip

\runningdefault{}

\subsection{Concolically Executing Default Terms}
\label{sec:default:concolic}

We now present how to support default terms during concolic execution.
We start with some preliminaries about concolic execution, and
the related concept of symbolic execution.

\paragraph{Background: Symbolic and Concolic Execution.}

Symbolic
execution~\cite{king1976symbolic,boyer75select,howden77symbolic,clarke76system}
is a program analysis technique that aims to explore all feasible program paths
in a program, and that has been successfully applied to a wide range of
languages~\cite{pasareanu2010symbolic,cadar08klee,micinski15symdroid,cadar08exe,DBLP:conf/uss/PoeplauF20,chipounov11s2e,david2016binsec}.
In symbolic execution, concrete inputs are replaced by symbolic values, and a
symbolic interpreter is tasked with executing the program under test. As the
program is executed, the interpreter will collect symbolic constraints
characterizing a given program execution path through a symbolic \emph{path
constraint}.  When hitting a conditional branching, the symbolic interpreter
will add the condition to the current path constraint, query an SMT
solver~\cite{z3,cvc5,barrett14problem,conchon:hal-01960203} to determine
whether the path constraint is satisfiable, and resume execution of the
program.  When a constraint is deemed unsatisfiable, the interpreter will
backtrack, and attempt to explore another execution path.

To make things more concrete, consider the small program $P \deq$
\mintinline[breaklines]{pascal}{if x > 0 then return 0 else if y < 10 then return 1 else error}.
We present the different steps of the symbolic execution in \Cref{fig:symexec}, showcasing
the generated path constraint tree at different steps.
Treating \texttt{x} and \texttt{y} as symbolic
variables, a symbolic interpreter would first reach the conditional branching
\mintinline{pascal}{if x > 0}, and generate two paths to explore, corresponding
to the path conditions $x > 0$ and $\falseconstr{x > 0}$
(\Cref{fig:sym-step1}).  As both path conditions are satisfiable, both
execution paths need to be explored.  Arbitrarily picking the first path
condition, it would then reach the terminal statement
\mintinline{pascal}{return 0}, terminating the execution of this path
(\Cref{fig:sym-step2}).  Executing the second path, it would then split the
path constraint tree again according to the constraint $y < 10$. As both path
conditions $\falseconstr{x > 0}, y < 10$ and $\falseconstr{x > 0},
\falseconstr{y < 10}$ are satisfiable, the symbolic interpreter would explore
both paths, thus detecting that the error statement is reachable
(\Cref{fig:sym-final}).

\begin{figure}[h]
  \vspace{-1em}
  \begin{subfigure}[T]{0.25\textwidth}
    \centering
    \begin{tikzpicture}[level distance=10mm,level/.style={sibling distance=18mm/#1}]
      \node {\mintinline{pascal}{if x > 0}}
      child {node {?} edge from parent node [left,yshift=3pt] {\scriptsize $x > 0$}}
      child {node {?} edge from parent node [right,yshift=3pt] {\scriptsize $\falseconstr{x > 0}$}};
    \end{tikzpicture}
    \caption{First step}
    \label{fig:sym-step1}
  \end{subfigure}
  \begin{subfigure}[T]{0.25\textwidth}
    \centering
    \begin{tikzpicture}[level distance=10mm,level/.style={sibling distance=18mm/#1}]
      \node {\mintinline{pascal}{if x > 0}}
      child {node {\mintinline{pascal}{return 0}} edge from parent node [left,yshift=3pt] {\scriptsize $x > 0$}}
      child {node {?} edge from parent node [right,yshift=3pt] {\scriptsize $\falseconstr{x > 0}$}};
    \end{tikzpicture}
    \caption{Second step}
    \label{fig:sym-step2}
  \end{subfigure}
  \begin{subfigure}[T]{0.55\textwidth}
    \centering
    \begin{tikzpicture}[level distance=10mm,level/.style={sibling distance=28mm/#1}]
      \node {\mintinline{pascal}{if x > 0}}
      child {node {\mintinline{pascal}{return 0}} edge from parent node [left,yshift=3pt] {\scriptsize $x > 0$}}
      child {node {\mintinline{c}{if y < 10}}
        child {node {\mintinline{pascal}{return 1}} edge from parent node [left,yshift=3pt] {\scriptsize $y < 10$}}
        child {node {\mintinline{pascal}{error}} edge from parent node [right,yshift=3pt] {\scriptsize $\falseconstr{y < 10}$}}
        edge from parent node [right,yshift=3pt] {\scriptsize $\falseconstr{x > 0}$}
      };
    \end{tikzpicture}
    \caption{Final state}
    \label{fig:sym-final}
  \end{subfigure}

  \vspace{1em}
  \caption{Path constraint trees at different stages of symbolic execution. Nodes marked as ?
    are not yet explored}
  \label{fig:symexec}
\end{figure}

Despite its successes, symbolic execution however faces several limitations.
Symbolic execution tools heavily rely on SMT solvers, and therefore
struggle when considering symbolic constraints beyond theories well-supported
by automated solvers, such as non-linear arithmetic. Additionally, symbolic
interpreters typically require access to the analyzed code, and hence do
not work well when interacting with an external environment or external library
calls~\cite{DBLP:journals/csur/BaldoniCDDF18,cadar13symbolic,godefroid07compositional}.

To address these issues, \emph{concolic execution}, a combination of
\emph{conc}rete and symb\emph{olic} execution was proposed
\cite{DBLP:conf/sigsoft/SenMA05,DBLP:conf/pldi/GodefroidKS05}. In concolic
execution, the program is executed with concrete inputs, and instrumented to
collect symbolic constraints during execution.
When one execution terminates,
some of the constraints in the path condition are negated and new inputs are
generated through a call to an SMT solver, leading to the exploration of a new
program path.

Compared to symbolic execution, concolic execution has several advantages.
First, when faced with unsupported logical theories, concolic execution can
still make progress by executing the program, instead of getting stuck when
encountering complex constraints. Second, by relying on concrete inputs covering
a wide range of program paths, concolic execution naturally provides a
reproducible testbed that can be used for automated unit testing, even when
symbolic execution would be imprecise due to complex program statements or calls
to libraries whose code is unavailable~\cite{godefroid07compositional}. Last,
while the generation of new inputs is typically performed by an automated solver
to guarantee the exploration of new program paths, it can also be combined with
other input generation techniques such as
fuzzing~\cite{stephens16driller,noller18badger,godefroid08automated,majumdar07hybrid}.

To formally describe concolic execution, we will rely on the following notations
for instrumented semantics. When $e \exevalstar e'$ denotes the concrete evaluation
from expression $e$ to expression $e'$, $\concolicglo{\rchi}{e} \exevalstar \concolicglo{\rchi'}{e'}$
will denote that, under initial path condition $\rchi$, $e$ evaluates to $e'$ with
new path condition $\rchi'$.

Coming back to our toy program $P$, concolic testing would therefore proceed as follows.
Let us assume that we start with inputs $x = 5, y = 5$.
Then, we would have $\concolicglo{\cdot}{P} \exevalstar \concolicglo{x > 0}{0}$, i.e.,
the concrete execution of the program returned the value 0, and the path condition
$x > 0$ was collected. A concolic engine would then negate part of the path condition,
query the SMT solver with the new constraints, i.e., $\falseconstr{x > 0}$,
and start another iteration with the inputs generated by the solver.
This process would repeat until negating constraints does not lead
to novel execution paths and all feasible program paths have been explored, with
concrete inputs leading to each of these paths.

\paragraph{Encoding Default Terms.}

To concolically execute default terms, we now need to provide a symbolic
representation to complement the concrete semantics presented in
\Cref{fig:default-reduction}.  To do so, we show how to instrument the concrete
semantics to collect symbolic constraints characterizing the current execution
path, by materializing the control flow structure of the evaluation of a
default expression. In order to cover all program paths induced by default
terms, this encoding must therefore consider all combinations of raised
exceptions, as well as a branching due to the default condition
$\synvar{e_{just}}$ when no exception is triggered.

We formally define our instrumented semantics in \Cref{fig:concolicdefault}.
We slightly extend the notations presented in the previous section, and denote
as \concolic{\rchi}{e}{C} a default expression $e$ with symbolic constraints
$\rchi$ and $C$. The constraint $\rchi$ corresponds to the path condition
collected up to the evaluation of the current default term.  The constraint
$C$, which we will call \emph{local constraint}, corresponds to the constraints
collected during the evaluation of the current term. The distinction between
$\rchi$ and $C$ is irrelevant for the rules presented in this section, and
$\concolic{\rchi}{e}{C}$ can safely be understood as $\concolicglo{\rchi \wedge
C}{e}$, i.e., the current path condition is the conjunction of $\rchi$ and $C$.
The importance of this separation will appear in \Cref{sec:default:optims},
when discussing several optimizations to concolic execution of default terms.
When either $\rchi$ or $C$ are trivial, we will omit them to simplify notations.

\begin{figure}[!t]
  {
  \text{Notations}
  \begin{align*}
    &\boxed{\concolic{\rchi}{e}{C}}&\text{\enquote{Expression $e$ has path condition $\rchi$ and local constraint $C$}}\\
    &\boxed{\concolicglo{\rchi}{e}}&\text{\enquote{Syntactic sugar for \concolic{\rchi}{e}{\cdot}}}\\
    &\boxed{\concolicloc{e}{C}}&\text{\enquote{Syntactic sugar for \concolic{\cdot}{e}{C}}}\\
    &\boxed{\synvar{e}}&\text{\enquote{Syntactic sugar for \concolic{\cdot}{e}{\cdot}}}\\
    &{\boxed{\concolic{\rchi}{e}{C} \exeval \concolic{\rchi'}{e'}{C'} }\hspace{-2em}}&\text{``Under path condition $\rchi$ and local constraint $C$,}\\[-.5em]
    &                                                                &\hspace{-5em}\text{ $e$ evaluates to $e'$ with new path condition $\rchi'$ and local constraint $C'$''}
  \end{align*}
  \centering
  \begin{mathpar}
    \inferrule[C-DefaultExpr]
    {e \exevalstar \concolicglo{\rchi'}{v}\quad\quad\quad\quad v \neq \synerror{}}
    {\concolic{\rchi}{\synlangle v_1 , \synpunct{\ldots,} v_i , e \synpunct{,\ldots} \synmid e_{just}  \synjust e_{cons} \synrangle}{C}
      \exeval
     \concolic{\rchi}{\synlangle v_1 , \synpunct{\ldots,} v_i, v \synpunct{,\ldots} \synmid e_{just} \synjust e_{cons} \synrangle}{C \wedge \rchi'}
    }
    \todo[inline]{PG I added v1 in C-DefaultExpr to be consistent}

    \inferrule[C-DefaultError]
    {e \exevalstar \concolicglo{\rchi'}{\synerror{}}}
    {\concolic{\rchi}{\synlangle v_1 \synellipsis v_i, e \synellipsis \synmid e_{just} \synjust e_{cons} \synrangle}{C} \exeval \concolicglo{\rchi \wedge \rchi' \wedge C}{\synerror{}}}

    \inferrule[C-DefaultTrueNoExceptions]
    {e_{just} \exevalstar \concolicglo{\rchi'}{\syntrue}}
    {\concolic{\rchi}{\synlangle \synemptydefault{}\synellipsis\synemptydefault{}\synmid e_{just} \synjust e_{cons} \synrangle}{C} \exeval \concolicglo{\rchi \wedge C \wedge \rchi' \wedge e_{just}}{e_{cons}}}

    \inferrule[C-DefaultFalseNoExceptions]
    {e_{just} \exevalstar \concolicglo{\rchi'}{\synfalse}}
    {\concolic{\rchi}{\synlangle \synemptydefault{}\synellipsis\synemptydefault{}\synmid e_{just} \synjust e_{cons} \synrangle}{C} \exeval
     \concolicglo{\rchi \wedge C \wedge \rchi' \wedge \falseconstr{e_{just}}}{\synemptydefault{}}
    }

    \inferrule[C-DefaultOneException]
    {v \neq \synemptydefault, \synerror}
    {\concolic{\rchi}{\synlangle \synemptydefault\synellipsis\synemptydefault\syncomma v \syncomma\synemptydefault\synellipsis\synemptydefault
      \synmid  e_{just} \synjust e_{cons}\synrangle}{C}
     \exeval \concolicglo{\rchi \wedge C}{v}}

    \inferrule[C-DefaultExceptionsConflict]
    {v_i \neq \synemptydefault \\ v_j \neq \synemptydefault \\ \forall k, v_k \neq \synerror}
    {\concolic{\rchi}{\synlangle v_1 \synellipsis v_i \synellipsis v_j \synellipsis v_n \synmid
      e_{just} \synjust e_{cons} \synrangle}{C} \exeval \concolicglo{\rchi \wedge C}{\synerror{}}}
  \end{mathpar}
  }
  \caption{Concolic semantics for default terms}
  \label{fig:concolicdefault}
\end{figure}

To illustrate how our semantics operates, we will rely on the following
example, where $x$ and $b$ are respectively integer and boolean variables:
\default{\pdefault{b}{1},\pdefault{x=0}{2}}{x>0}{3}.  Let us assume that we
want to concolically execute this term with $x = 3$ and $b = \concreteTrue$.
Initially, both the global constraints $\rchi$ and local constraints $C$
are trivial.
Following rule \textsc{C-DefaultExpr}, we must first reduce the exception $\pdefault{b}{1}$ to a value.
Since we have a default term (cf. \Cref{not:defaults}) and its default condition $b$ holds, we can apply
rule \textsc{C-DefaultTrueNoExceptions}, obtaining the concolic term \concolicglo{b}{1}.
We conclude the application of rule \textsc{C-DefaultExpr}, leading to the concolic term
\concolicloc{\default{1,\pdefault{x=0}{2}}{x>0}{3}}{b}.
We then perform a similar reduction on the second exception by combining rules \textsc{C-DefaultExpr} and \textsc{C-DefaultFalseNoExceptions} to obtain the concolic term \concolicloc{\default{1, \synemptydefault}{x > 0}{3}}{b \wedge \falseconstr{x = 0}}.
As exactly one exception reduced to a non-empty or error value, the final step in the
concolic execution is then to apply rule \textsc{C-DefaultOneException}; local constraints
are thus appended to the current (empty) path condition.
The execution therefore terminates
with the value \concolicglo{b \wedge \falseconstr{x = 0}}{1}, accurately capturing that any
set of inputs satisfying the path constraint $b \wedge \falseconstr{x = 0}$ will follow the
same execution path.

\todolater{AF: I removed mentions of soundness/completeness, they feel more confusing at this stage, and they are hinted at with the sentence "accurately capturing ... same execution path"}

\paragraph{A Complete Concolic Execution.}

Equipped with our concolic semantics, we can now define a concolic execution
operating on default terms. To do so, we follow the standard workflow of
concolic execution: starting from an initial, arbitrary set of inputs, we
concolically execute the program, and collect the corresponding path
constraints. Once the execution terminates, we choose another unexplored path,
querying the SMT solver to generate an input satisfying the new path
constraint, and therefore leading to a different execution. We repeat this
process until all paths have been explored, or an exploration timeout has been
reached.

To illustrate how a complete concolic execution operates, we will reuse the default term \default{\pdefault{b}{1},\pdefault{x=0}{2}}{x > 0}{3}
previously presented. Starting from inputs $x = 3$ and $b = \concreteTrue$, we previously saw that this term reduced
to the value $1$, generating the constraints $b$ and $\falseconstr{x = 0}$. By negating the last constraint and querying the solver,
we obtain new inputs $x = 0$ and $b=\concreteTrue$, and perform another concolic iteration. Repeating this process
leads to the generation of 5 different testcases fully covering the program, and identifying two execution paths
leading to $\synerror$ and $\synemptydefault$ respectively. The summary of the concolic execution is available in \Cref{fig:steps}.

\begin{figure}[h]
    \centering
    \begin{tabular}{c c c c l r}
      \toprule
        Step & \synvar b & \synvar x & Output & Constraints after evaluation & Next constraints to try \\
        \midrule
        \# 1 & \concreteTrue & \concrete 3 & \concrete 1
          & $[\trueconstr{b}, \falseconstr{x=0}]$
          & $b \land x=0$
            \\
        \# 2 & \concreteTrue & \concrete 0 & \concrete \synconflict
          & $[\trueconstr{b}, \trueconstr{x=0}]$
          & $\lnot b$
            \\
        \# 3 & \concreteFalse & \concrete {3} & \concrete 3
          & $[\falseconstr{b}, \falseconstr{x=0}, \trueconstr{x>0}]$
          & $\lnot b \land \lnot (x=0) \land \lnot (x>0)$
            \\
        \# 4 & \concreteFalse & \concrete {-1} & \concrete\synempty
          & $[\falseconstr{b}, \falseconstr{x=0}, \falseconstr{x>0}]$
          & $\lnot b \land x=0$ \\
        \# 5 & \concreteFalse & \concrete 0 & \concrete 2
          & $[\falseconstr{b}, \trueconstr{x=0}]$
                                                                             & --\\
      \bottomrule\\
    \end{tabular}
  \vspace{-1em}
  \caption{Concolic testcase generation for $\default{\pdefault{b}{1},\pdefault{x=0}{2}}{x > 0}{3}$}
  \label{fig:steps}
\end{figure}

\section{\name: Implementing Concolic Execution}
\label{sec:overview}

Relying on the default logic and its concolic interpretation presented in the previous section,
we now present \name, a concolic execution engine for the Catala programming language.

\begin{figure}[!t]
  \begin{multicols}{2}
\begin{minted}{catala_en}
```catala
declaration structure Household:
 data income content money
 data nb_children content integer

declaration scope IncomeTaxComputation:
 input house content Household
 internal tax_rate content decimal
 output income_tax content money
```

## Article 1
The income tax for an individual is
defined as a fixed percentage of the
individual's income over a year.
```catala
scope IncomeTaxComputation:
 definition income_tax equals
   house.income * tax_rate
```

## Article 2
The fixed percentage mentioned at
article 1 is equal to 20%.
```catala
scope IncomeTaxComputation:
  definition tax_rate equals 20%
```

## Article 3
If the individual's income is less
than $10,000, the fixed percentage
mentioned at article 1 is equal to 10%.
```catala
scope IncomeTaxComputation:
 exception definition tax_rate
  under condition house.income <= $10,000
  consequence equals 10%
```

## Article 4
If the individual is in charge of 3 or
more children, then the fixed percentage
mentioned at article 1 is equal to 15%.
```catala
scope IncomeTaxComputation:
 exception definition tax_rate
  under condition house.nb_children >= 3
  consequence equals 15%
```
\end{minted}
  \end{multicols}
  \vspace{-1em}
  \caption{Running example: a simplified income tax computation}
  \label{fig:ex:running}
\end{figure}

Catala is a recent domain-specific language tailored to implement computational
laws, relying on default logic under the hood~\cite{merigoux2021catala}.  We
show an example Catala program in \Cref{fig:ex:running}, encompassing the
default term for the simplified income tax computation described in
\Cref{sec:defaultterms}. One important aspect of Catala is its literate
programming capabilities, allowing to reflect the structure of the law in the
implementation. As seen in this example, each unit of law is immediately
followed by its implementation. The main purpose is to facilitate the
maintainability and transparency of the implementation with respect to the law
test; at compilation time, Catala will rely on the Markdown code markers
\mintinline{catala_en}{```catala} to extract the implementation.  We refer the
interested reader to the Catala tutorial~\cite{catalatutorial} for a more
detailed description of the merits of literate programming when implementing
computational law.

We now provide an overview of the Catala language on our running example.
Catala requires strongly
typed declarations exposing the interfaces of a program.  Here, lines 2-4
define a \texttt{Household} record, defined by an \texttt{income} and a number
of children.  Scopes are the basic abstraction unit in Catala; they can be
conceived as a loose equivalent to functions in other programming languages.  A
scope is first statically declared, with a typed declaration of its inputs,
outputs and internal variables.  In our example, lines 6-9 define the interface
of our income tax computation, which computes an income tax for a given
household.  The main case of the income tax computation is defined lines 17-19 and 26-27:
the default tax rate is 20\% of the household's income. This rate is amended
by two cases, one setting it at 10\% for low incomes (lines 35-38),
the second setting the rate to 15\% for large families (lines 46-49).
Note that while these new
definitions are following one another in this simplified example,
real-world Catala implementations might have amended these definitions in several
different files, for instance if the base case is defined in the Tax Code while the
large families exception is defined in the Family Code. During compilation,
the Catala toolchain will combine the three definitions of \texttt{tax\_rate} in
\texttt{IncomeTaxComputation} through the use of a default term, corresponding
to the one defined in the previous section: \runningdefault{}

\begin{figure}[!t]
  \centering
  \begin{tikzpicture}[node distance=.7cm and .7cm]
    \tikzstyle{block} = [scale=0.8,rectangle, draw,
    text width=5em, text centered, minimum height=3em]
    \tikzstyle{label} = [midway,scale=0.7,rectangle,
    text width=6em, text centered]

    \node[block] (src) {Source code};
    \node[right=of src] (scopelang) {$\ldots$};
    \node[block, right=of scopelang] (dcalc) {Default calculus};
    \node[block, right=of dcalc] (scalc) {Statement calculus};

    \node[block, below =1em of dcalc, xshift=1.2cm] (interpreter) {Interpreter};
    \node[block, below =1em of dcalc, fill=green!20, xshift=-1.2cm] (cutecat) {\name};

    \node[right=of scalc,yshift=.5cm] (python) {\includegraphics[width=.5cm]{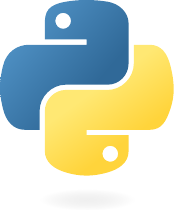}};
    \node[right= of scalc,yshift=-.5cm] (c) {\includegraphics[width=.5cm]{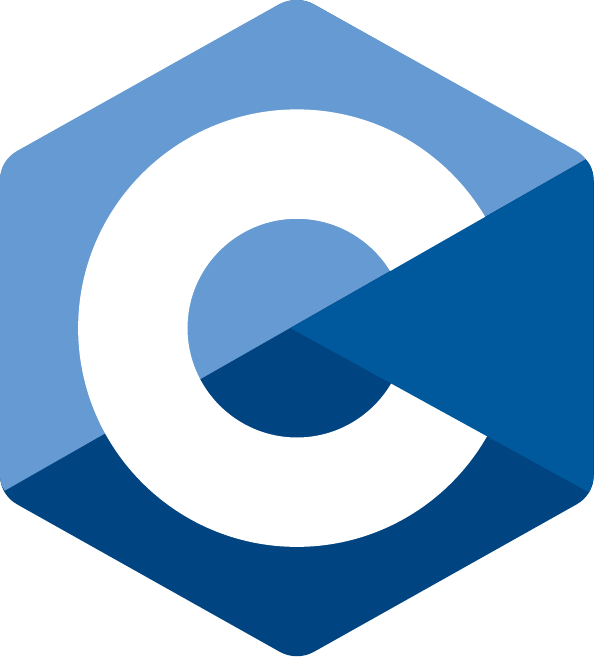}};

    \draw[->] (src) -- (scopelang);
    \draw[->] (scopelang) -- (dcalc);
    \draw[->] (dcalc) -- (scalc);
    \draw[->] (dcalc) -- (interpreter);
    \draw[->] (dcalc) -- (cutecat);
    \draw[->] (scalc) -- (python);
    \draw[->] (scalc) -- (c);
  \end{tikzpicture}
  \caption{Passes of the Catala compiler, and \name integration}
  \label{fig:catala:passes}
\end{figure}

Our concolic engine, \name, has been integrated directly into the Catala
toolchain, whose relevant parts are shown in \Cref{fig:catala:passes}. To allow
integration of Catala implementations in existing projects, the compiler
translates Catala source code into mainstream programming languages like C or
Python. To do so, it relies on a series of intermediate representations (IRs),
elided here, until reaching the default calculus. The default calculus is one of
the last IRs in the compilation pipeline, and is very close to the statement
calculus which transforms expressions into statements before emitting C or
Python code.

Default terms are explicitly materialized in the default calculus; it is where
the reference Catala interpreter operates, following the semantics for default
terms described in \Cref{fig:default-reduction}. This is therefore a natural
choice for the implementation of our concolic engine; additionally, being
located at the same compilation step of the interpreter nullifies any
discrepancies of potential bugs introduced earlier in the compilation pipeline.
We implemented our concolic interpreter as a fork of the standard interpreter,
instrumenting it to collect symbolic constraints. This design allows us to
closely follow the reference semantics of Catala, thus minimizing implementation
mistakes, while also simplifying the maintenance of \name when the Catala
language evolves.

Similarly to other works on concolic execution~\cite{DBLP:conf/sigsoft/SenMA05,majumdar07hybrid},
\name relies on a depth-first search (DFS) strategy to explore new execution paths.
In a DFS exploration, a concolic engine generates new inputs by negating the last
constraint added to the path condition.
Concretely, during a concolic iteration, \name keeps the constraints $C$ used
to generate the current input. Each constraint is annotated to indicate whether
it has already been negated. The execution generates a path condition $C'$,
which is compared to $C$; all constraints in $C' \setminus C$ are marked as new, and
the last new constraint in $C'$ is negated to generate a new input.  This
workflow corresponds to a DFS exploration of the path constraint tree; however,
it only requires keeping the last explored branch in memory instead of the
entire tree, therefore improving the performance of concolic execution.

To illustrate how \name{} operates, we now describe the concolic execution
of our running example. \name{} generates here four different cases,
in an order determined by the values used in the first iteration; in our case,
they are set to $\texttt{income} = \$0 $ and $\texttt{nb\_children} = 0$.
\begin{enumerate}
\item \texttt{income} is \$0 and there are no children. In that case, only the first exception (lines 35-38 of \Cref{fig:ex:running}) can be applied, and the tax is \$0.
\item \texttt{income} is \$0 and there are three children. In that case, both exceptions (lines 35-38 and 46-49) can be applied.
  A conflict is raised to the user, showing that the law has been unfaithfully translated, or that it is ambiguous.
\item \texttt{income} is \$10,000.01 and there are two children. In that case, the general case (lines 26-27) applies, and the tax is \$2,000 (monetary amounts are rounded to the cent).
\item \texttt{income} is \$10,000.01 and there are three children. In that case, the exceptional case at lines 46-49 applies and the tax is \$1,500.
\end{enumerate}

By default, \name outputs these different cases directly to the user, however,
we also provide facilities to directly generate Catala \emph{test scopes}.
These test scopes can be saved, and easily replayed using the Catala toolchain,
either by using the reference interpreter or by compiling
\todolater{PG try to compile generated testcases? For now I interpret to see my overhead, but the execution time of compiled testcases may also be a good metric?}
it to one of the Catala backends.

As identified by our concolic execution, our running example can raise
conflicts, i.e., when considering large, low-income households.  This conflict
is not due to an implementation error, but rather to an ambiguity in our
(simplified) law. Resolving this ambiguity requires legal interpretation, which
must be performed by lawyers, and not programmers. In this case, legal
precedents might suggest that ambiguities shall be resolved in the most
favorable way to citizens, therefore defining a 10\% tax rate. To implement
this decision, programmers could thus modify the Catala implementation of
Article 3 as shown in \Cref{fig:ex:fixed}, by materializing a priority order on the
evaluation of both exceptions: as the exception is now defined on \texttt{income\_tax},
this exception will trigger first, before reaching the base computation depending on
\texttt{tax\_rate}\footnote{%
    Note that a more idiomatic fix would make use of Catala "labels", that
    enable programmers to order exceptions explicitly. We omit this solution to
    keep our presentation of Catala small. The Catala tutorial \cite{catalatutorial}
    has more information on labels.}.

\begin{figure}[!t]
  \centering
\begin{minted}{catala_en}
scope IncomeTaxComputation:
 exception definition income_tax
  under condition house.income <= $10,000
  consequence equals house.income * 10%
\end{minted}
  \vspace{-1em}
  \caption{Removing conflicts in income tax computation}
  \label{fig:ex:fixed}
\end{figure}

\name{} is implemented in 3,400 lines of OCaml code, and relies
on the Z3 SMT solver~\cite{z3}, called through its OCaml bindings.
It is an integral part of the Catala toolchain, and
therefore leverages its build system, simplifying the use of \name{}
on Catala projects. To provide users with more control on the concolic
execution, \name{} allows the use of assertions in the source code to restrict
the input space, for instance, to specify that only households in a given
country or state must be considered.
Users can provide hints about initial inputs for the concolic execution,
in order to quickly explore variants of a given situation.
\todolater{PG implement that? R: It can just be CLI for default values per type I think}

\paragraph{Divisions and Rounding.}

In addition to our handling of default terms described in \Cref{sec:encodingdefault},
\name{} supports the concolic execution of most Catala constructs, including
arbitrary-precision integers, booleans, money and decimal expressions; structures and field operations; algebraic data types and pattern-matching;
functions and function calls; and conditionals. The concolic execution of most of these
operations is standard, and we therefore omit their presentation. Two points are
however of particular interest, namely, arithmetic divisions and rounding.

Divisions by zero are a well-known source of runtime errors. To reason about them,
concolic tools therefore consider divisions as implicitly branching,
and collect whether the denominator is equal to zero as a symbolic constraint.\footnote{
Note that our semantics conflates runtime errors with conflicts, and returns
$\synerror$ in both cases.
Indeed, the conflict value acts as a \emph{de facto}
uncatchable exception: when it appears, it stops the evaluation of the program,
which immediately returns.
From the user's perspective, the main concern is whether the program can lead
to an error, independently of which one.
As we generate concrete inputs, it is always possible to run the reference
Catala interpreter to provide precise error reporting.
}

By providing different representations for numerical values, i.e.,
decimal (represented as rationals) and integers values, Catala also
requires conversions between these types; when casting a decimal
to an integer, this is implemented as a rounding operator.
Catala's rounding convention is to round to the nearest integer; when
two are equidistant, it returns to the one furthest away from 0.
To model this semantics, we define a custom $round$ Z3 function,
defined as \mintinline{pascal}{round(q) = if q>=0 then floor(q + 1/2) else -floor(-q + 1/2)};
the \texttt{floor} operation is implemented using Z3's native casting from rationals to integers.

As part of our experimental evaluation of \name{} (\Cref{sec:evalexp}), our precise
modeling of rounding led us to find inputs where the different Catala backends
were inconsistent: rounding operations in Catala's Python backend
exhibited minor differences with the reference interpreter. We reported this
issue to the Catala developers, and upstreamed a fix to the compiler.

\paragraph{Limitations.}
\name{} currently supports a large subset of Catala which is sufficient to
analyze real-world programs (\Cref{sec:evalexp}) that mainly rely on
integer-rational reasoning as well as algebraic datatypes.
However, operations on lists and dates are only partially
supported. List operations in Catala include filtering depending on a
condition, aggregating over list contents to compute all sorts of values, and
applying a function to all elements in the list; providing symbolic encoding of
such operations would require higher-order symbolic
reasoning~\cite{tobin12higher,nguyen15relatively}.
On the other hand, dates can easily be represented in an SMT solver as the number
of days since a specific point in time, e.g., the Unix epoch. This representation
allows to model several operations, such as the addition of a (possibly symbolic)
number of days, or the duration corresponding to the difference between two dates.
Unfortunately, other operations, including month and year addition or returning
the first day of a week or a month, have a more complex
semantics~\cite{DBLP:conf/esop/MonatFM24} which does not naturally fit into
this SMT encoding.

To handle these operations, we leverage the strengths of concolic execution
over symbolic execution, and do not generate any symbolic encoding, relying
instead entirely on concrete evaluation at the cost of completeness.
We leave the exploration of suitable symbolic models for these operations to
future work.

Finally, our implementation currently only supports the Z3 SMT solver, as it
directly provides ready-to-use OCaml bindings. This limitation could be easily
lifted, as we can already generate SMT-LIB files~\cite{BarFT-SMTLIB} for each
SMT query, which can be consumed by most other SMT
solvers~\cite{cvc5,barrett14problem,conchon:hal-01960203}. Additionally, our
prototype is single-threaded; the recent introduction of Multicore
OCaml~\cite{sivaramakrishnan20retrofitting} nevertheless paves the way for
parallelizing concolic execution~\cite{bucur11parallel,staats10parallel}.

\section{Improving the Scalability and Usability of \name}
\label{sec:optims}

To improve the scalability of \name, we now discuss various optimizations and
heuristics to enhance the performance of concolic execution, as well as the
usability of our toolchain by non-expert users.  The concrete impact of these
optimizations will be evaluated experimentally in \Cref{sec:evalexp}.

\subsection{Optimizing Concolic Execution of Default Terms}
\label{sec:default:optims}

The approach presented in \Cref{sec:default:concolic} allows us to perform concolic execution
on default terms. However, to improve interactivity with developers, our goal is not only to
exhaustively analyze all program execution paths, but also to find possible errors
as early as possible during the analysis. In this section, we therefore propose
several optimizations aiming to prune constraints leading to redundant cases, and to
prioritize executions that might lead to conflict errors.

\paragraph{Lazily Evaluating Exceptions.}

When evaluating a default term, the semantics presented in \Cref{fig:default-reduction}
requires first evaluating all exceptions, except if one of them raises a conflict error
(\textsc{DefaultExpr}). If two exceptions evaluate to a non-empty value, the ensuing
conflict is therefore only detected when applying rule \textsc{DefaultExceptionsConflict}.
As concrete executions are in practice very fast, this has little performance
impact when concretely executing a default term; furthermore, from a usability perspective,
it is helpful to report all conflicting cases to users.
However, the overhead becomes larger during concolic execution, where we must explore all possible execution
paths when evaluating remaining exceptions, including many calls to an SMT
solver.

To circumvent this issue, we propose in \Cref{fig:default-lazy} alternative semantic rules
to replace \textsc{DefaultExpr} and \textsc{DefaultExceptionsConflict}, as well as their concolic counterparts.
These new rules stop the evaluation as soon as two exceptions return a non-empty
value. This prevents exploring the remaining exceptions; the multiple execution
paths would all lead to a conflict in the current default term.

Consider for instance a default term $e := \default{e_1, e_2, e_3}{e_{just}}{e_{cons}}$, where
$e_1$ and $e_2$ both evaluate to values that are neither empty nor a conflict, respectively
generating constraints $C_1$ and $C_2$. By following rule \textsc{C-DefaultExpr}, concolically
executing $e$ would require concolically executing $e_3$ and generating its associated constraint
$C_3$, which would then be added to the path condition when applying rule \textsc{C-DefaultExceptionsConflict}.
Further iterations of concolic execution would therefore negate the last constraints in the path condition,
i.e., those in $C_3$, while preserving $C_1$ and $C_2$. As any input satisfying $C_1 \wedge C_2$ leads
to $e_1$ and $e_2$ evaluating to non-empty nor conflict values, thus raising a conflict, this approach
would induce a number of redundant iterations that is exponential in the number of constraints in $C_3$.
By instead adopting rule \textsc{C-DefaultExceptionsConflict-Lazy} and stopping the execution after the evaluation
of $e_2$, we therefore prune $C_3$ from the path condition, thus reducing the number of iterations needed.

\begin{figure}[!t]
  \centering
  \smaller
  \begin{mathpar}
    \inferrule[DefaultExpr-Lazy]
    {e \exevalstar v\quad\quad\quad\quad v \neq \synerror{}}
    {\synlangle \synemptydefault \synellipsis \synemptydefault , e \synellipsis \synmid e_{just}  \synjust e_{cons} \synrangle
      \exeval \synlangle \synemptydefault \synellipsis \synemptydefault, v \synellipsis \synmid e_{just} \synjust e_{cons} \synrangle}

    \inferrule[DefaultExprOne-Lazy]
    {e \exevalstar \synemptydefault{}\quad\quad\quad v_i \neq \synerror,\synemptydefault}
    {\synlangle \synemptydefault \synellipsis v_i \synellipsis \synemptydefault{} , e \synellipsis \synmid e_{just}  \synjust e_{cons} \synrangle
      \exeval \synlangle \synemptydefault \synellipsis v_i \synellipsis \synemptydefault{} , \synemptydefault \synellipsis \synmid e_{just}  \synjust e_{cons} \synrangle }

    \inferrule[DefaultExceptionsConflict-Lazy]
    {e \exevalstar v\quad\quad\quad\quad v \neq \synerror{},\synemptydefault\quad\quad\quad v_i \neq \synerror,\synemptydefault}
    {\synlangle \synemptydefault \synellipsis v_i \synellipsis \synemptydefault{} , e \synellipsis \synmid e_{just}  \synjust e_{cons} \synrangle \exeval \synerror }

    \inferrule[C-DefaultExpr-Lazy]
    {e \exevalstar \concolicglo{\rchi'}{v}\quad\quad\quad\quad v \neq \synerror{}}
    {\concolic{\rchi}{\synlangle \synemptydefault \synellipsis \synemptydefault , e \synpunct{,\ldots} \synmid e_{just}  \synjust e_{cons} \synrangle}{C}
      \exeval
     \concolic{\rchi}{\synlangle \synemptydefault \synellipsis \synemptydefault, v \synpunct{,\ldots} \synmid e_{just} \synjust e_{cons} \synrangle}{C \wedge \rchi'}
    }

    \inferrule[C-DefaultExprOne-Lazy]
    {e \exevalstar \concolicglo{\rchi'}{\synemptydefault}\quad\quad\quad\quad v_i \neq \synerror,\synemptydefault}
    {\concolic{\rchi}{\synlangle \synemptydefault \synellipsis v_i \synellipsis\synemptydefault , e \synpunct{,\ldots} \synmid e_{just}  \synjust e_{cons} \synrangle}{C}
      \exeval
     \concolic{\rchi}{\synlangle \synemptydefault \synellipsis v_i \synellipsis\synemptydefault, \synemptydefault \synpunct{,\ldots} \synmid e_{just} \synjust e_{cons} \synrangle}{C \wedge \rchi'}
    }

    \inferrule[C-DefaultExceptionsConflict-Lazy]
    {e \exevalstar \concolicglo{\rchi'}{v}\quad\quad\quad\quad v \neq \synerror,\synemptydefault\quad\quad\quad\quad v_i \neq \synerror,\synemptydefault}
    {\concolic{\rchi}{\synlangle \synemptydefault \synellipsis v_i \synellipsis \synemptydefault, e \synpunct{,\ldots} \synmid e_{just}  \synjust e_{cons} \synrangle}{C}
      \exeval
     \concolicglo{\rchi \wedge \rchi' \wedge C}{ \synerror{} }
    }

  \end{mathpar}
  \caption{Alternative concrete and concolic lazy semantics for default terms}
  \label{fig:default-lazy}
\end{figure}

\paragraph{Reorganizing Exceptions.}

By relying on the semantics presented in \Cref{fig:default-lazy}, concolic executions
of default terms can therefore be greatly shortened if exceptions evaluating
to non-empty values are at the front of the exception list. Beyond not evaluating
additional exceptions, their corresponding symbolic constraints will not be
added to the path condition, thus reducing the size of the constraint tree
and the number of iterations needed for concolic execution to terminate.

To leverage this fact, one key observation is that the result of the evaluation
of a default term is independent of the evaluation order of the exception list.
We formalize this property in \Cref{thm:default-swap}, which states that
swapping any two exceptions results in evaluating the default term to the
same value. %
This therefore allows to use different
evaluation orders depending on the program and current symbolic state. In our
implementation, which we describe in more detail in the next sections, we
permute exceptions to group them depending on their free variables, a technique
reminiscent of constraint reordering introduced by \citet{cadar08klee}.

\begin{theorem}[Independence of the exception evaluation order]
If there exists a default value $v$ such that
$\synlangle \synpunct{\ldots,} e_i \synellipsis e_j \synpunct{,\ldots} \synmid e_{just} \synjust e_{cons} \synrangle \exevalstar v$,
then
$\synlangle \synpunct{\ldots,} e_j \synellipsis e_i \synpunct{,\ldots} \synmid e_{just} \synjust e_{cons} \synrangle \exevalstar v$
\label{thm:default-swap}
\end{theorem}
\todolater[inline]{PG proof sketch or lemma? If we have room...}

\paragraph{Pruning Non-Conflict Path Conditions.}
As a last optimization, we present in \Cref{fig:default-conflict-prune}
an alternate formulation of rule \textsc{C-DefaultError}, which applies
when evaluating an exception yields a conflict value.
If $\concolicglo{\rchi}{\default{e_1 \synellipsis e_i, e, \synpunct{\ldots}}{e_{just}}{\allowbreak e_{cons}}}
  \exevalstar
  \concolic{\rchi}{\default{v_1 \synellipsis v_i, e, \synpunct{\ldots}}{e_{just}}{e_{cons}}}{C}$,
and $e \exevalstar \concolicglo{\rchi'}{\synerror}$, then the application of rule
\textsc{C-DefaultError} will add both $C$ and $\rchi'$ to the path condition.
We observe however that this will lead to redundant executions, as any input satisfying
$\rchi'$ will lead to $e$ evaluating to $\synerror$. Pruning $C$ from the path condition
thus avoids unneeded concolic exploration, which we formally capture through rule
\textsc{C-DefaultError-Early}.

This rule directly derives from \Cref{thm:default-swap}; indeed,
the rule \textsc{C-DefaultError-Early} is conceptually equivalent to reorganizing
the exception list to place $e$ at the head of the list, and then applying
\textsc{C-DefaultError}. This is where the need for differentiating between the
path condition $\rchi$ and local constraint $C$ arises: while constraints
corresponding to the evaluation of earlier exceptions can be safely dropped,
the path condition up to this default term must be preserved.

\begin{figure}[!t]
  \centering
  \begin{mathpar}
    \inferrule[C-DefaultError-Early]
    {e \exevalstar \concolicglo{\rchi'}{\synerror{}}}
    {\concolic{\rchi}{\synlangle v_1 \synellipsis v_i, e \synellipsis \synmid e_{just} \synjust e_{cons} \synrangle}{C} \exeval \concolicglo{\rchi \wedge \rchi'}{\synerror{}}}
  \end{mathpar}

  \caption{Alternative concolic semantics for conflict-inducing evaluations}
  \label{fig:default-conflict-prune}
\end{figure}

\paragraph{Implementation.}\label{sec:opt:default}
\name{} implements the lazy evaluation of exceptions, as well as exception
reorganization to order exceptions according to their sets of free variables.
Pruning non-conflict path conditions is trickier to implement while following
a DFS exploration strategy.
A key invariant of DFS exploration is that, during a concolic iteration,
the previous path condition (used to generate the current input)
is a prefix of the current path condition up to the negated constraint.
This does not hold with rule \textsc{C-DefaultError-Early},
as the local constraint $C$ is dropped from the path condition.
We leave the study of alternative exploration strategies and their impact
as future work.

\subsection{Generating Human-compatible Testcases}

While many testcases might be equivalent from a semantic perspective (i.e.,
following the same program path), some might be easier to review by lawyers and
humans in general.  Consider for instance the constraint $\falseconstr{x \leq
\$10,000}$, used in our running example to determine whether a household is in
the low-income category. As we have seen in \Cref{sec:overview}, Z3 might
generate $x = \$10,000.01$ as a testcase.
When crafting testcases, lawyers would instead lean towards round numbers,
e.g., $x = \$11,000$, which would be easier to use when redoing manual
computations of legal statutes to compare them to the implementation's output.
To mimick this behavior, \name{} therefore provides heuristics attempting
to generate semantically equivalent, ``human-friendlier'' testcases.

To do so, we rely on the encoding of optional constraints in the
solver, dubbed ``soft constraints''~\cite{z3softconstraints}.
Z3 natively allows to combine SMT reasoning with solving optimization
objectives, including the specification of soft constraints; however, we
encountered significant slowdowns when attempting to use this feature, raising
scalability issues, that have been reported to Z3 developers.
We posit that this is due to non-linearity in our soft-constraints -- e.g.,
when specifying $x \% 100 = 0$ -- and in our rounding function defined in \Cref{sec:overview}.
We instead adopt an alternative approach, by implementing custom soft
constraints through multiple queries to the solver. Concretely, when an
iteration of concolic execution terminates and a new path condition $C$ is
generated, we first query Z3 to check whether $C$ is satisfiable. If so, we
further query Z3 by iteratively refining $C$ with additional constraints
forcing human-friendlier input generation.

Our current soft constraints particularly target monetary inputs. We first
attempt to force such inputs to be multiples of 100. If impossible, we then try
with multiples of 10, and finally with integers, aiming to prevent the use of
cents. Empirically, these constraints have a significant impact: as monetary
amounts are encoded at the cent level, Z3 frequently returns instances with
cents, which are less readable and less pleasant to use when computing by hand,
especially when percentages and rounding are involved. This paves the way for
further improvements to \name's usability; beyond monetary inputs, we envision
the development of a lawyer-friendly custom configuration language, enabling
the specification of preferred soft constraints depending on needs and
usecases.

\subsection{Pattern-Matching Case Folding}
\label{sec:pm}

The Catala language supports simple pattern matching on the constructors of
algebraic datatypes.  By default, \name{}'s interpretation of pattern matching
considers each case of the pattern matching as a different branch.
That way, the logic of the pattern matching is materialized in the
constraint tree itself, and the engine is tasked with successively trying each
branch. However, we have noticed that
this approach can be highly inefficient in existing Catala codebases.

Consider for example the snippet in \cref{fig:ex:pattern}, extracted from the
Catala implementation of the French housing benefits. At lines 5-13, this
program performs a match on the 9 variants of a sum type representing
geographic areas of France (mainland, or overseas territories). To analyze
this program, a naive implementation of our concolic interpreter would therefore
consider 9 different, disjoint branches, leading to 9 distinct iterations.

Here however, the arms of several cases are identical; patterns
can be conjoined to only create 2 branches, one corresponding to
cases returning \mintinline{pascal}{true}, the other to cases returning
\mintinline{pascal}{false}. Doing so heavily reduces the size of the path constraint
tree; in more realistic examples, performance gains can be significant when
such patterns appear deep inside expressions, requiring long concolic executions.
We implemented a case folding optimization for pattern matching
inside \name, folding similar cases by considering the disjunction of their patterns.

\begin{figure}[!t]
  \begin{multicols}{2}
\begin{minted}{catala_en}
scope HousingBenefitsForRentals:
  exception definition family_rate
  under condition
    (match area with
     -- Guadeloupe: true
     -- Martinique: true
     -- LaRéunion: true
     -- Mayotte: true
     -- SaintBarthélemy: true
     -- SaintMartin: true
     -- Guyane: false
     -- Métropole: false
     -- SaintPierreEtMiquelon: false)
  consequence equals ...
\end{minted}
  \end{multicols}
  \vspace{-1em}
  \caption{French housing benefits case where pattern case folding is beneficial}
  \label{fig:ex:pattern}
\end{figure}

\subsection{SMT-Solving Optimizations}
\label{sec:opt:smt}

\paragraph{Trivial constraint simplification.}
The path constraints taken into account during the concolic execution can
sometimes be trivial.  Indeed, some conditions do not depend on input
variables, or are compilation artifacts from the translation of the source code
into the default calculus, where tautological conditions are sometimes added in
default expressions. Negating these trivially true
constraints in the hope of exploring other execution paths is pointless,
as the new constraint given to the solver will be
trivially false, and therefore unsatisfiable.

To avoid this, we query the SMT solver to simplify constraints before they
are added in the path, and then remove trivial constraints statically.  This
avoids spurious calls to the SMT solver.  Additionally, we perform a linear
check for trivial unsatisfiability, by checking whether the newest constraint
is a negation of a previous one. This last technique is well known by
the concolic solver community, and already mentioned in the seminal work of
\citet{DBLP:conf/sigsoft/SenMA05}.

\paragraph{Incremental SMT solving.}
Z3 provides incremental solving capabilities, where sets of constraints can be
incrementally added or removed through stack-based \texttt{push} and
\texttt{pop} operators. When popping the stack, the solver backtracks to an
earlier proof state, only removing lemmas learned between \texttt{push} and \texttt{pop}
and avoiding reproving many facts.
We implemented support for incremental solving, pushing a new stack for each
element in the path condition. As \name{}'s exploration strategy is currently
fixed to depth-first search (DFS), incremental solving is particularly appealing:
constraints are naturally changed in a last-in-first-out fashion, corresponding
to a stack-based data structure.

\section{Evaluating \name}
\label{sec:evalexp}

We now turn to our experimental evaluation of \name. We first
quantify the impact of the optimizations described in \Cref{sec:optims},
relying on four Catala projects predating this work (\Cref{sec:ablation}).
We then discuss \name{}'s ability to identify conflicts (\Cref{sec:mutation}),
before evaluating \name{} on the largest Catala program up to this day, namely,
an implementation of the French Housing Benefits (\Cref{sec:housing}).
All experiments have been performed on a desktop machine featuring an Intel
Core i7-12700 and 128GB of DDR5 RAM (although we have never seen CUTECat use
more than 1GB), running Ubuntu 24.04.1, OCaml 4.14.1 and Z3 4.12.5.

\subsection{Evaluating \name's Optimizations}
\label{sec:ablation}

\paragraph{Benchmarks Overview.}
To evaluate the performance of \name and its optimizations, we have selected
four different, real-world Catala programs as benchmarks for our evaluation.
These programs predate our implementation of \name, and have been written
by third-party, experienced Catala developers.  One benchmark corresponds to
the implementation of a fragment of Section 132 of the US Tax Code, defining
qualified employee discounts.  Other benchmarks implement
diverse bodies of law in France: one codifies minimum salary (SMIC), another
the computation of a family quotient used in the income tax computation (Family
Quotient).  The last benchmark computes the housing
benefits in the rental case, where additional restrictions have been made:
households are located in mainland France and contain less than 10 children.
These benchmarks reflect multiple facets of the law, and therefore exhibit
diverse encodings in Catala; we provide quantitative details about them in
\Cref{fig:benchmarks}, namely, the lines of Catala code (including legal
specification), the number of branching constructs they contain (default terms,
if-then-else, pattern matches), and the size of the compiled Python code. We
particularly observe that the default terms outnumber the other branching
terms, arguing in favor of their importance.

\paragraph{Evaluation Metrics.}
To evaluate the completeness of concolic tools, a standard approach is to
measure the code coverage achieved. In our context, this is however hard to
perform: the Catala interpreter does not provide any coverage measurements,
while results on compiled Catala code, e.g., in Python, are highly unreliable.
Indeed, on a small program for which we can manually determine all possible
execution paths, a single Python execution reaches a line coverage of 65\%,
while executing all program paths barely reaches 80\%. We suspect that this is
due to a bloated translation of default logic into mainstream languages.
\todolater[inline]{R: Mention that \name is able to terminate and exhaustively explore all the benchmarks?}
Our evaluation therefore focuses on analysis times, and on the number of testcases
generated. We defer a discussion of the completeness of \name{} to \Cref{sec:mutation}.

\begin{table}[!t]
  \centering
  \begin{tabular}{lr@{\hskip 1em}r@{\hskip 1em}r@{\hskip 1em}r@{\hskip 1em}}
    \toprule
    Benchmark           &   \href{https://github.com/pierregoutagny/catala-examples/blob/174aac026e46cc5cfaa657fed79cfb6aa88b746c/us_tax_code/section_132.catala_en}{\textcolor{blue}{Section 132}} &   \href{https://github.com/pierregoutagny/catala-examples/blob/174aac026e46cc5cfaa657fed79cfb6aa88b746c/smic/smic.catala_fr}{\textcolor{blue}{SMIC}} &  \href{https://github.com/pierregoutagny/catala-examples/tree/174aac026e46cc5cfaa657fed79cfb6aa88b746c/impot_revenu}{\textcolor{blue}{Family Quotient}} & \href{https://github.com/pierregoutagny/catala-examples/tree/174aac026e46cc5cfaa657fed79cfb6aa88b746c/aides_logement}{\textcolor{blue}{Housing Benefits}} \\
\midrule
 Catala LOC          &           133 &    368 &               776 &              19655 \\
 \quad nb(default terms)   &            15 &     17 &                46 &                251 \\
 \quad nb(if-then-else)    &             3 &      0 &                24 &                171 \\
 \quad nb(match)           &             2 &      0 &                14 &                139 \\
 Python LOC          &           343 &    586 &              1413 &              31826 \\
    \bottomrule\\[-.5em]
  \end{tabular}
  \caption{Statistics about the benchmarks used in the ablation study}
  \label{fig:benchmarks}
  \centering
  \begin{tabular}{llr@{\hskip 1em}r@{\hskip 1em}r@{\hskip 1em}r@{\hskip 1em}r@{\hskip 1em}r@{\hskip 1em}r@{\hskip 1em}r}
    \toprule
      Category      &   \multicolumn{2}{c}{Section 132} &   \multicolumn{2}{c}{SMIC} &   \multicolumn{2}{c}{Family Quotient} &   \multicolumn{2}{c}{Housing Benefits} \\
                        \cmidrule(r){2-3} \cmidrule(r){4-5} \cmidrule(r){6-7} \cmidrule(r){8-9}
      Optimizations     & None & All      & None & All      & None & All      & None & All\\
    \midrule
      Solver Calls    & 41 & 24  & 138 & 138 & 13224 & 4142  & 355848  & 126147  \\
 Generated Tests & 10 & 10 & 17  & 17  & 381    & 381    & 24995   & 17435 \\
    \bottomrule\\[-.5em]
  \end{tabular}
  \caption{Impact of optimizations on solver calls and generated tests}
  \label{fig:steps_and_tests}
  \vspace*{-2em}
\end{table}

\paragraph{Solver Calls and Generated Tests.}

We compare in \Cref{fig:steps_and_tests} the number of solver calls performed
and the number of tests generated by \name{}.  These numbers are provided both
for naive executions without any optimization enabled and when all
optimizations are enabled.  The difference in generated tests evaluates the
impact of pruning redundant exploration paths through pattern matching case
folding; the difference in solver calls represents the additional impact of
our optimizations targeting the SMT encoding, i.e., our handling of
trivial constraints.  While these optimizations lead to
improvements on almost all benchmarks (SMIC has no matches or trivial
constraints, so it is not affected), their precise impact heavily differs.
Avoiding calls to solver when considering trivial constraints
is almost always beneficial; it reduces the number of calls up to 3.2x for the
family quotient benchmark.  Conversely, pattern matching folding only helps the
housing benefits; other, smaller examples do not rely as much on large pattern
matching.
The tables below provide a more detailed analysis of the impact of individual
optimizations.

\paragraph{Incremental Solving.}

We evaluate in \Cref{fig:incr} the gains provided by Z3's incremental mode
(\Cref{sec:opt:smt}).  To do so, we compare both the total execution times and
the solver execution times when running \name.  Similarly to other
works~\cite{DBLP:conf/sigsoft/SenMA05}, our empirical results suggest that
incremental solving is a cornerstone of efficient concolic execution; it
provides more than an order of magnitude improvement to the solving times.  On
the smaller examples, these order-of-magnitude improvements are also observed
on the total running time of the concolic engine, as the SMT solving dominates
the running times.  On larger programs, e.g., housing benefits, \name{}'s
execution time is however dominated by running the concolic interpreter;
solving constraints only represents 35\% of the total time.
On this example, incremental solving is 32x faster than Z3's default mode,
leading to a total gain of more than 35 minutes.

\begin{table}[!t]
  \centering
  \begin{tabular}{l@{\hskip 1em}r@{\hskip 1em}r@{\hskip 1em}r@{\hskip 1em}r@{\hskip 1em}r@{\hskip 1em}r@{\hskip 1em}r@{\hskip 1em}r}
    \toprule
    Options                 & \multicolumn{2}{c}{Section 132}  & \multicolumn{2}{c}{SMIC}    & \multicolumn{2}{c}{Family Quotient} & \multicolumn{2}{c}{Housing Benefits} \\                \cmidrule(r){2-3} \cmidrule(r){4-5} \cmidrule(r){6-7} \cmidrule(r){8-9}
                            & Total & Solve                    & Total & Solve               & Total & Solve                       & Total & Solve\\
    \midrule
    Standard                & 0.27s & 0.23s                    & 1.01s & 0.85s               & 82.61s & 69.46s                    & 6002.03s & 2262.38s \\
    Incremental             & 0.02s & 0.01s                    & 0.08s & 0.01s               & 5.21s  & 0.23s                      & 4306.72s & 69.88s \\
    \bottomrule\\[-.5em]
  \end{tabular}
  \caption{Solver and total execution times, with and without incremental solving.}
  \label{fig:incr}
  \vspace*{-2em}
\end{table}

\paragraph{Ablation Study.}

To evaluate the impact of each evaluation, we perform an ablation study and compare total running times for each optimization individually
against the bare implementation of \name{} (without any optimization), and all optimizations enabled at once.
Results are presented in \Cref{fig:times}.
We consider the following optimizations:
lazy-default and exception packing (\Cref{sec:opt:default}),
incremental solving (\Cref{sec:opt:smt}),
constraint simplifications (\Cref{sec:opt:smt}),
pattern matching case folding (\Cref{sec:pm}),
and frontend optimizations already offered by the Catala compiler. They currently consist in partial evaluation of booleans,
  and simplification of trivial defaults, i.e., when there are no exceptions and the default condition is a boolean value.

We observe that almost all optimizations provide noticeable performance improvements, but with variability on
benchmarks likely tied to variance in their structure. For smaller programs where solving time dominates,
the largest improvements are due to incremental solving.

\begin{figure}[!t]
  \centering
  \includegraphics[width=0.78\textwidth]{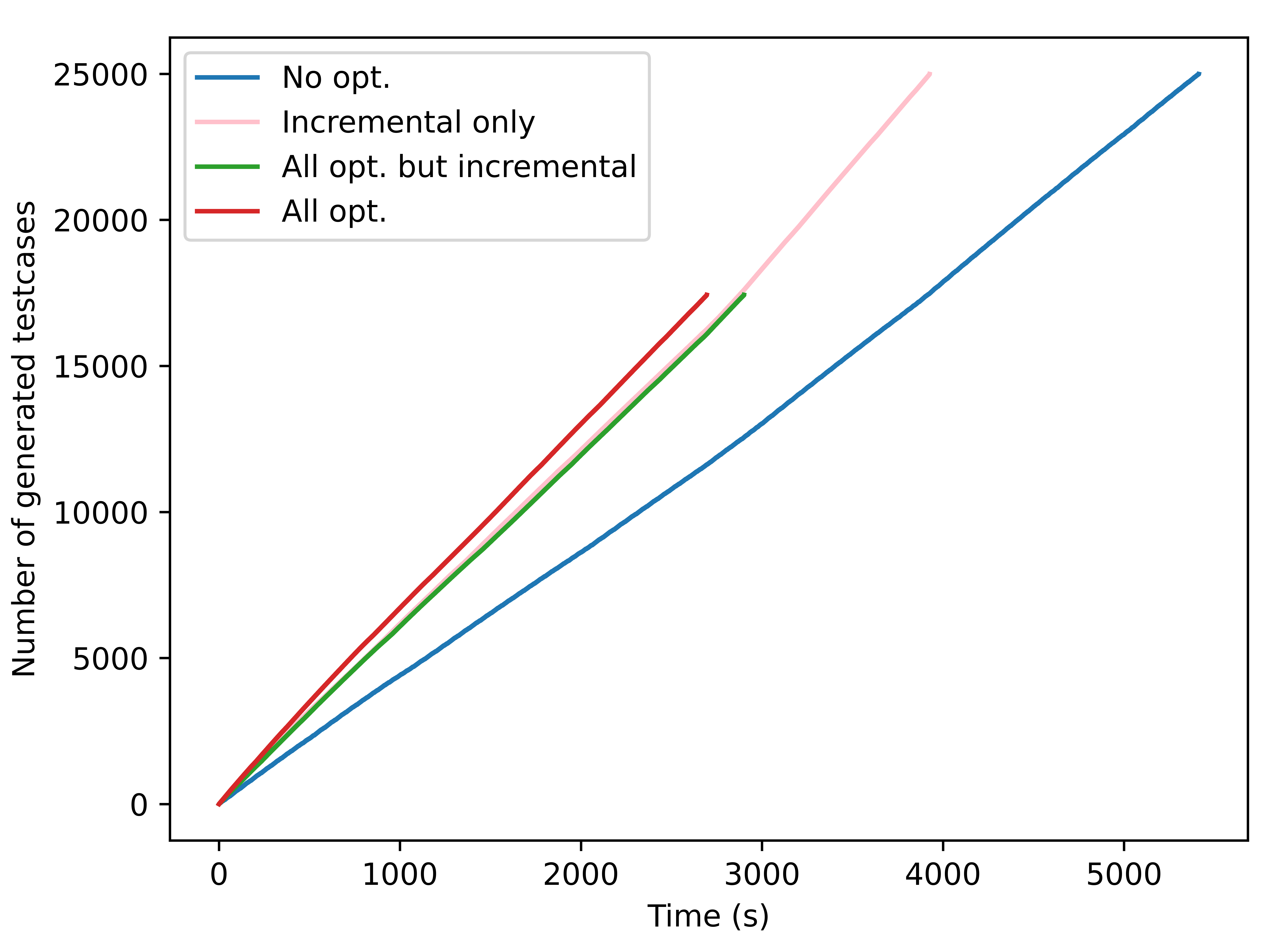}
  \vspace{-1em}
  \caption{Comparison of test generation rates on the housing benefits benchmark}
  \label{fig:testrate}
\end{figure}

For housing benefits however, other optimizations reduce the total time by an additional 25\%.
To better describe this example, we measure in \Cref{fig:testrate} the evolution of the
number of tests generated through time. Findings include several results previously observed:
the optimizations prune the exploration tree by removing redundant cases, and incremental
solving significantly improves total execution time. Additionally, we observe
a linear generation rate, suggesting that concolic execution does not struggle to reach
certain paths, which could have been a shortcoming of less systematic approaches, e.g., black-box fuzzing.

Notably, optimizations related to default terms, i.e., lazy default and exception packing do not impact
the running times of \name{}. This is however expected: these optimizations are only of interest when
terms can return conflict values. Our benchmarks only contain mature Catala programs, corresponding
to implementations of enacted laws. To the best of our knowledge, no corpus of conflict-inducing Catala examples
currently exists, making a precise evaluation of these optimizations difficult. We however expect these
optimizations to shine when applied during development processes, as well as to identify possible
inconsistencies during the preparation of new legislation.

\todolater{Some optimizations affect others: trivial is not that interesting in an incremental context? AF: Just by ablation, hard to evaluate. We would need the combination of incr + trivial.
I skipped this description}

\begin{table}[!t]
  \centering
  \begin{tabular}{l@{\hskip 1em}r@{\hskip 1em}r@{\hskip 1em}r@{\hskip 1em}rr}
    \toprule
    Optimizations & Section 132    & SMIC         & Family Quotient   & \multicolumn{2}{c}{Housing Benefits}\\
    \midrule
    None              & 0.27s ± 0.00  & 1.01s ± 0.01 & 82.61s ± 0.55     & 6002.03s ± 19.46   \\
    Lazy-default      & 0.27s ± 0.00  & 1.02s ± 0.01 & 84.31s ± 1.15     & 5995.01s ± 15.21   \\
    Exception packing & 0.27s ± 0.00  & 0.98s ± 0.01 & 83.41s ± 1.58     & 6013.71s ± 22.12   \\
    Incremental       & 0.02s ± 0.00  & 0.08s ± 0.00 & 5.21s ± 0.04      & 4306.72s ± 77.78   \\
    Trivial           & 0.09s ± 0.00  & 0.99s ± 0.01 & 12.62s ± 0.11     & 4258.93s ± 22.88   \\
    Match folding     & 0.21s ± 0.00  & 0.99s ± 0.01 & 46.89s ± 0.54     & 4045.87s ± 21.30   \\
    Frontend opts.    & 0.17s ± 0.00  & 0.99s ± 0.00 & 38.20s ± 0.42     & 3538.32s ± 22.37   \\
    All               & 0.02s ± 0.00  & 0.08s ± 0.00 & 4.34s ± 0.07      & 2843.01s ± 15.02   \\
    \midrule
    Speedup           & 1250.00\%      & 1162.50\%     & 1803.46\%          & 111.12\%            \\
    \bottomrule\\[-.5em]
  \end{tabular}
  \caption{Ablation study for \name optimizations. Average on eight runs, with standard deviation displayed after ±.}
  \label{fig:times}
  \vspace{-3em}
\end{table}

\paragraph{Soft Constraints.}
We evaluate the impact of using soft constraints in \Cref{fig:soft}; the evaluation
is performed with all optimizations enabled. We only consider the Section 132 and housing benefits
benchmarks, as our soft constraints target monetary amounts
given as inputs, which other examples do not have.
On large, representative Catala examples, we observe that the computational
overhead of using soft constraints is small. Additionally, the soft constraints
are satisfied for almost all cases, significantly improving the usability of
\name{}. Indeed, 87.04\% of the 17435 generated testcases can be solved with a
soft constraint ensuring monetary amounts are multiples of 100, and additionally
9.98\% and 2.80\% for the multiples of 10 and units, respectively. In total,
only 30/17435 generated testcases (0.17\%) do not satisfy any soft
constraint. Our encoding of soft constraints particularly benefits from
incremental solving: they are only added to a path condition proved satisfiable
by the solver -- which corresponds to refining an existing solution.
Furthermore, incremental solving significantly reduces solving time, minimizing
the impact on \name's running time of calling the solver several times on a
given path. As such, we expect further extensions to soft constraints to have a
minor impact on \name's performance.

\todolater{add option for script}

\begin{table}[!t]
  \centering
  \begin{tabular}{l@{\hskip 1em}rHH@{\hskip 1em}r}
    \toprule
    Benchmark    & Section 132   & SMIC         & Family Quotient   & Housing Benefits   \\
    \midrule
    All         & 0.02s ± 0.00  & 0.08s ± 0.00 & 4.34s ± 0.07      & 2843.01s ± 15.02   \\
    Soft. cons. & 0.03s ± 0.00  & 0.07s ± 0.00 & 3.98s ± 0.05      & 2968.80s ± 16.90   \\
    \midrule
    Slowdown    & 50.00\%        & -12.50\%      & -8.29\%            & 4.42\%              \\
    \midrule
    Number of tests & 10         &          17  &              381  &  17435\\
    \% sat soft cons. (multiples of 100)      & 100\%     &          0\% &               0\% &     87.04\%\\
    \% sat soft cons. (multiples of 10)      & N.A.     &          0\% &                0\% &      9.98\%\\
    \% sat soft cons. (multiples of 1)      & N.A.     &          0\% &                0\% &      2.80\%\\
    \bottomrule\\[-.5em]
  \end{tabular}
  \caption{Impact of soft constraints}
  \label{fig:soft}
  \vspace*{-2em}
\end{table}

\subsection{Detecting Errors with \name}
\label{sec:mutation}

We now aim to determine whether \name{} is effective in detecting issues in computational law implementations.
Unfortunately, existing Catala codebases are mature and implementing real-world, enacted laws;
they are therefore highly unlikely to exhibit such issues.
To evaluate \name's completeness, we therefore turn to mutation testing~\cite{DBLP:journals/tse/JiaH11}
to automatically inject errors in Catala programs, thereby simulating coding mistakes or legislative imprecisions.

We apply mutation testing to our most complex example: housing benefits. To do so, we first
randomly select a default term in the program; we then either remove a chosen number of
exceptions (possibly leading to unhandled cases), duplicate an exception (thus creating
a conflict), or negate the default condition of a default term (possibly leading to
unhandled cases). We follow this process to generate 20 new programs, which we manually
inspect to confirm that mutations introduced issues. In particular, we must ensure
that the default term considered is reachable, and that assumptions and conditions
in other parts of the code do not preclude the unhandled cases, or the execution of the conflicting exceptions.

\name{} successfully identifies all 20 issues, providing concrete inputs to identify problematic cases.
On average, \name{} requires 0.55 seconds to reach the injected bug; this execution
time leads us to believe that our tool would be particularly suitable as part of continuous integration
workflows, but also to evaluate intended amendments to laws.

\subsection{Case Study: Rental Housing Benefits}
\label{sec:housing}

Our evaluation so far focused on a constrained case of the rental housing benefits,
considering only households located in mainland France and with less than 10 children.
We chose to enforce these restrictions due to the multiple executions required to evaluate
different \name{} settings, which already required more than one hour each.
In this section, we now explore the general, unconstrained case of the rental housing benefits.
With all optimizations enabled, \name{} is able to analyze this program in 6h37m, generating 186390 testcases.
This process generates 1338575 solver calls; the total solver time is 366s.

Interestingly, \name's analysis led to the discovery of a conflict error in the existing implementation.
The housing benefits are defined for various cases, including when the flat is shared with other roommates, or when a single bedroom is rented.
However, an interpretation conflict can happen for people claiming rental housing benefits if they are sharing a bedroom with roommates.
This conflict was previously manually discovered by \citet{merigoux:hal-03933574} during their implementation of the housing benefits.
Despite querying relevant public administrations, this case was deemed unlikely to happen, and no legal interpretation was
provided to resolve this ambiguity. No other errors have been found by \name{} on the rental housing benefits.

A manual inspection of the generated testcases revealed that the current input space for the housing benefits is under-constrained.
For example, some testcases correspond to households located overseas and in housing zone 3 -- but to the best of our knowledge,
all overseas territories are defined to be in housing zone 2.
Such cases exemplify where expert knowledge from lawyers would be beneficial:
provided with indications about which inputs are well-formed, programmers can then
restrict \name{}'s search space through program assertions, therefore speeding up
the analysis.

\paragraph{Overhead of Concolic Execution.}
We now compare the overhead of the concolic execution of \name{} with the
interpretation times of all the generated testcases. The reference Catala
interpreter evaluates the 186390 generated tests in 1h29, meaning \name{} has a
4.5x overhead. While large, this overhead is not unexpected: the concolic
interpreter instruments each expression evaluation to collect symbolic
constraints, and performs repeated calls to the SMT solver. Prior work on
concolic execution observed similar trends: for example,
\citet%
      {DBLP:conf/uss/Yun0XJK18} measure that KLEE \cite{cadar08klee}
has up to a three order-of-magnitude overhead, while our experimental results
are in line with approaches focusing on reducing this
overhead~\cite{DBLP:conf/uss/PoeplauF20,DBLP:conf/uss/ChenHYZSLYS22}.
This leads us to believe that \name's scalability is comparable to concolic
tools in other languages; we nevertheless intend to perform a deeper profiling
of our interpreter to identify potential bottlenecks.
This comparison also allows us to sanity-check our implementation with respect
to Catala's semantics: we confirmed that \name and the Catala reference
interpreter agreed on all generated testcases. Furthermore, we can use those
same testcases to check whether all Catala backends have the same behavior on
the case study.

\section{Related Work}

\paragraph{Concolic Execution.}
Concolic execution was introduced by the seminal works of
\citet{DBLP:conf/pldi/GodefroidKS05} and \citet{DBLP:conf/sigsoft/SenMA05}
almost two decades ago.  Since then, it has been applied to various programming
languages~\cite{cadar08exe,cadar08klee,DBLP:conf/cav/SenA06,luckow2016jdart,DBLP:conf/ecoop/MarquesS0A22}.
We refer the reader to the
survey of \citet{DBLP:journals/csur/BaldoniCDDF18} for an extensive coverage of
concolic and symbolic execution.  As Catala has
different features compared to traditional, imperative programming languages,
most of the advanced techniques mentioned in the survey (memory encoding, loop
summarization) are not relevant for our work.  Similarly, human-readability of
generated tests is usually not an objective of traditional concolic execution
engines.  However, \citet{DBLP:journals/csur/BaldoniCDDF18} mention that ``the
way high-level switch statements are compiled can significantly affect the
performance of path exploration'', echoing in an imperative setting the need
for an encoding similar to the pattern-matching case folding presented in
\Cref{sec:pm}.  \citet{DBLP:journals/scp/GiantsiosPS17} support pattern
matching through a compilation to a decision tree, which is the approach used
to generate machine code \cite{DBLP:conf/ml/Maranget08}.  In \Cref{sec:pm}, we
chose to implement a simpler approach as Catala currently supports matching
against a single sum type.  As we have mentioned in \Cref{sec:mutation}, our
mutation approach is lacking information to be reliable in
user-constrained cases, which required us to manually inspect mutated
programs. To create an analysis-oriented benchmarking suite for Catala programs,
an interesting avenue would be to rely on the
program generation techniques of \citet{DBLP:conf/icse/VikramPS21}.

\paragraph{Formal Methods for the Law.}
Several approaches were previously proposed to reason about default logic.
\citet{risch1996analytic} and \citet{cassano2019tableaux} both define tableaux-based
approaches~\cite{howson2005logic} to reason about propositional formulas
expressed in default logic, while \citet{schaub1995new} and \citet{linke1995lemma}
propose query answering methodologies for default logic propositions.
By focusing on the underlying logic, these works differ from our approach tailored
to program verification, which emphasizes testcase generation and
requires reasoning about numerical expressions and program constructs such as
structures or enumerations.
In their compiler for the French tax code,
\citet{DBLP:conf/cc/MerigouxMP21} have relied on coverage-guided fuzzing with
AFL~\cite{afl} to improve the quality of a pre-existing test-suite. Their approach
however only generated 275 minimized testcases from a first generation of
30,000 cases. In contrast, \name's concolic-based approach allows to
generate cases exploring different execution paths, while reaching rare
ambiguous or unhandled situations.
\citet{DBLP:conf/esop/MonatFM24} and \citet{DBLP:conf/cpp/BorgesBRRBJ24}  focus
on formalizing date-duration arithmetic and semantics of UTC time respectively,
with the former also analyzing date-related ambiguities
in legal computations.
Our work currently only provides limited support for date operations,
but could build upon these formalizations to define suitable symbolic
encodings.
\todolater{I had added a small paragraph about LLMs, commented out due to space constraints}

\section{Conclusion and Future Work}

Legal expert systems implementing computational laws are pervasive,
and faithfully translating the law into code is error-prone.
To alleviate this issue, we proposed \name, a concolic execution tool
targeting computational law implementations in the Catala language.
\name{} relies on a novel concolic semantics for default terms, and
is equipped with several optimizations improving its scalability
and usability by lawyers.

\name scales to Catala codebases implementing real-world French and US laws, generating hundreds of thousands of unique testcases in less than seven hours of CPU time.
We believe this paper takes a step towards the use of formal methods during legislative processes.
Our future work includes tailoring \name's pipeline and interface for lawyers through interdisciplinary experimentation.
We also plan to discuss additional usecases for \name with law experts; namely, we believe \name could be used to improve
state-of-the-art computer-assisted tax teaching techniques~\cite{lawsky2021teaching}.

\subsubsection{Acknowledgments.}
We thank the anonymous reviewers for their constructive feedback and support of our work.
We are grateful to Rohan Padhye for his suggestions around concolic testing for Catala, and to Patrick Baillot for feedback about earlier versions of this paper.
We thank Liane Huttner \& Sarah Lawsky for the interesting discussions around the properties this work targets, Louis Gesbert for his technical help around the Catala compiler, and Denis Merigoux for clarifications about the current implementation of default terms in Catala.
We appreciated the many discussions and valuable feedback about this work we got from the whole Catala team.

\bibliography{conferences,cited}
\doclicenseThis

\end{document}